\documentclass[english,aps,pra,superscriptaddress,floatfix,notitlepage,reprint,pdftex,unicode=true,colorlinks=true,citecolor=Blue,linkcolor=RubineRed,urlcolor=Blue]{revtex4-2}
\usepackage{graphicx}
\usepackage{mathrsfs}
\usepackage{bm}
\usepackage{amsmath}
\usepackage{dcolumn}
\usepackage{epstopdf}
\usepackage{dsfont}
\usepackage{amssymb}
\usepackage{tabularx}
\usepackage{array}
\usepackage{float}
\usepackage{color}
\usepackage{epstopdf}
\usepackage{mathrsfs}
 \usepackage{booktabs}
\usepackage[colorlinks, linkcolor=blue,anchorcolor=blue,citecolor=blue,urlcolor=blue]{hyperref}
\newcommand{\tr}{\mathrm{Tr}}

 \usepackage{amsmath}

\newcommand{\ket}[1]{|#1\rangle}
\newcommand{\bra}[1]{\langle #1|}

\newcommand{\n}{\nonumber\\}

\newcommand{\up}{\uparrow}

\newcommand{\ex}[1]{\langle #1\rangle}

\begin{document}

\title{Optimal energy storage and collective charging speedup in the central-spin quantum battery}
\author{Hui-Yu Yang}
    \affiliation{School of Physics, Northwest University, Xi'an 710127, China}
\author{Kun Zhang}
    \affiliation{School of Physics, Northwest University, Xi'an 710127, China}
    \affiliation{Shaanxi Key Laboratory for Theoretical Physics Frontiers, Xi'an 710127, China}
    \affiliation{Peng Huanwu Center for Fundamental Theory, Xi'an 710127, China}
    
    \author{Xiao-Hui Wang}
    \email{xhwang@nwu.edu.cn}
    \affiliation{School of Physics, Northwest University, Xi'an 710127, China}
    \affiliation{Shaanxi Key Laboratory for Theoretical Physics Frontiers, Xi'an 710127, China}
    \affiliation{Peng Huanwu Center for Fundamental Theory, Xi'an 710127, China}

  \author{Hai-Long Shi}
 \email{hailong.shi@ino.cnr.it} 
\affiliation{QSTAR, INO-CNR, and LENS, Largo Enrico Fermi 2, 50125 Firenze, Italy}

\begin{abstract}
Quantum batteries (QBs) exploit principles of quantum mechanics to accelerate the charging process and aim to achieve optimal energy storage. 
However, analytical results for investigating these problems remain lacking due to the challenges associated with nonequilibrium dynamics.
In this work, we analytically investigate a central-spin QB model in which $N_b$ spin-1/2 battery cells interact with  $N_c$ spin-1/2 charger units, using  $m$ initially excited charger units as a resource.
By employing the invariant subspace method and the shifted Holstein-Primakoff  transformation, we identify four scenarios in which optimal energy storage can be achieved: 
(i) $N_b\!\ll\!m\!\ll\!N_c$; (ii) $m\!\ll\!N_b\!\ll\!N_c$; (iii) $m\!\ll\!N_c\!\ll\!N_b$; and (iv) $N_b\!\ll\!m\!=\!kN_c$ [$k\!\in\!(0,1)$].
In these cases, optimal storage is ensured by the SU(2) symmetry emerging from the charging dynamics.
The first three cases map the central-spin QB to different Tavis-Cummings (TC) QBs, while the fourth corresponds to the non-TC limit.
We analytically determine the charging time and demonstrate that in the fully charging cases (i) and (iv), the collective charging exhibits an $N_b$-fold enhancement in speedup compared to the parallel charging scheme.
Additionally, we numerically observe a unified charging behavior when $m\!=\!N_c$, showing that asymptotically optimal energy storage is possible when $N_b\!=\!m\!=\!N_c$. 
In this case, we find a collective charging enhancement scaling as $N_b^{0.8264}$.
The origin of the collective charging advantage in central-spin quantum batteries is also  analyzed through  the quantum speed limit and a multipartite entanglement witness.
Our results highlight the crucial role of dynamically emergent SU(2) symmetry in providing an analytical understanding of non-equilibrium charging dynamics in QBs.
\end{abstract}
    
\maketitle

\begin{figure*}[t]
        \includegraphics[width=2\columnwidth]{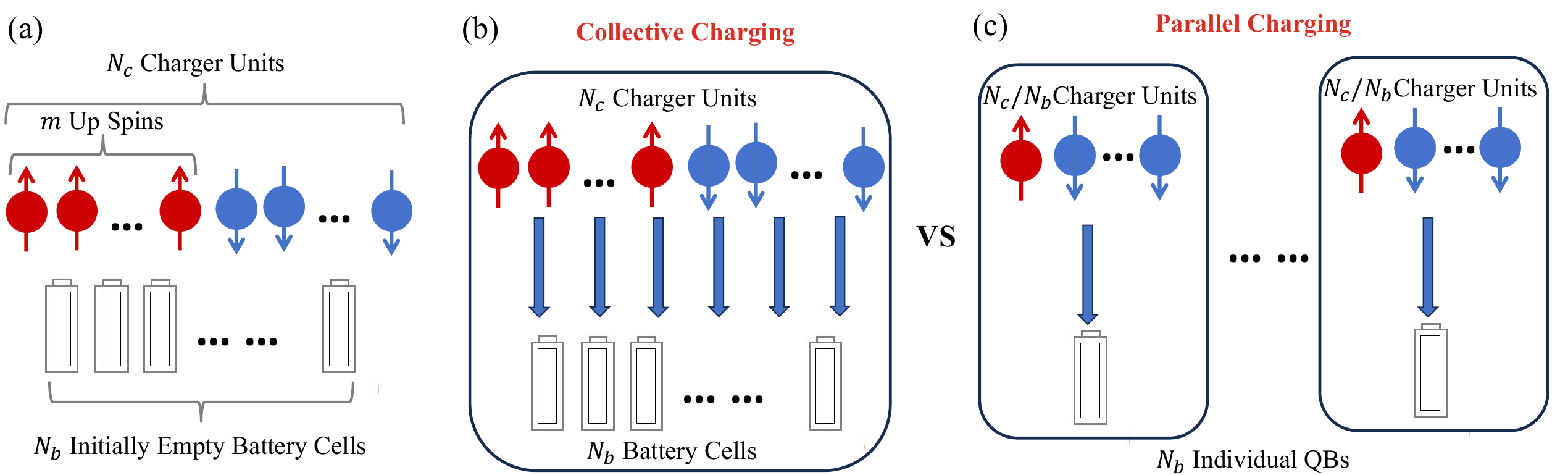}
         \caption{Schematic diagram of the central-spin QB. 
         (a)  Initially, the system consists of $N_b$ battery cells in their ground states, with no excitations, and $N_c$ charger units prepared in a Dicke state containing $m$ excitations.
Also shown is a comparison between (b) the collective charging scheme and (c) the parallel charging scheme.}\label{fig0}
\end{figure*}

\section{Introduction}
Pioneering theoretical work on energy extraction from many-body states has suggested that the generation of multipartite entanglement can enhance the efficiency of work extraction~\cite{PhysRevLett.111.240401}. This finding implies the existence of a new class of energy storage devices, known as quantum batteries~\cite{PhysRevE.87.042123} (QBs), which exploit unique quantum features to accelerate the charging process and outperform classical batteries.
Compared to charging individual battery cells independently, collective charging exhibits a significant quadratic speedup in charging time~\cite{PhysRevLett.128.140501,PhysRevLett.118.150601,PhysRevResearch.2.023113,gyhm2024beneficial}, 
which is believed to arise from multipartite entanglement induced by collective interactions.
The phenomenon where charging power scales faster than the number of quantum cells is referred to as collective charging  advantage~\cite{RevModPhys.96.031001}, which has been observed in various models, including the central-spin model~\cite{PhysRevA.103.052220}, spin-chain model~\cite{PhysRevA.97.022106,PhysRevE.104.024129}, Sachdev–Ye–Kitaev model~\cite{PhysRevLett.125.236402}, and Dicke model~\cite{zhang2023enhanced,PhysRevLett.120.117702,PhysRevB.105.115405}.

On the other hand, energy storage is another crucial metric for evaluating the performance of a quantum battery~\cite{PhysRevLett.129.130602,PhysRevA.109.012204,lu2024topological}. 
Practically, a QB that charges rapidly but stores only minimal energy is not desirable. 
Recent studies have explored the relationship between energy storage and quantum correlations~\cite{PhysRevLett.129.130602,PhysRevLett.122.047702,PhysRevLett.130.210401}. However, identifying many-body models that satisfy the conditions for achieving optimal energy storage remains challenging.
Here optimal energy storage refers to a situation where the battery absorbs all the energy initially stored in the charger or reaches a fully charged state, which will be rigorously defined later. 
Our recent work has shown that the emergence of SU(2) symmetry in the Tavis-Cummings (TC) model guarantees the realization of optimal energy storage~\cite{PhysRevA.109.012204}.
Could this emergent symmetry also explain the presence of optimal energy storage in other models?
Moreover, under conditions of optimal energy storage, can we achieve collective charging speedup simultaneously? 
We aim to address these questions in an important low-dimensional model.

The central-spin model has provided significant insights into the quantitative understanding of decoherence in nitrogen-vacancy (NV) centers in diamond, as well as the dynamics of entanglement~\cite{schliemann2003electron,PhysRevLett.96.140604,hanson2008coherent,PhysRevB.82.161308,PhysRevLett.109.140403,PhysRevB.101.184307,wan2020decoherence,PhysRevLett.121.080401,PhysRevB.101.155145,PhysRevB.99.174308,PhysRevB.88.155305,PhysRevA.72.052113,PhysRevLett.100.160505,PhysRevLett.132.250401,PhysRevA.75.032333}. 
This model can be naturally put into the framework of QBs, referred to as the central-spin QB, where $N_b$ central spins act as battery cells and $N_c$ bath spins serve as charging units.
It has been argued that a collective charging advantage is present in this model; however, the analysis is based on applying the Holstein-Primakoff (HP) transformation to both the bath and central spins~\cite{PhysRevA.103.052220}, which is valid only in the short-time regime and does not extend to the time scale including maximal energy storage~\cite{PhysRevLett.122.047702}.
On the other hand, the use of the HP approximation simplifies the central-spin model to the TC model, which we refer to as the TC limit. 
As discussed in Ref.~\cite{PhysRevA.109.012204}, even within the TC model, optimal energy storage is not guaranteed in all cases. 
In a specific non-TC limit~\cite{PhysRevB.104.245418}, we identified an optimal energy storage scenario for the case of $N_b\!=\!2$ and conjectured, based on numerical evidence, that optimal energy storage persists under this condition for the arbitrary $N_b$ case~\cite{PhysRevB.104.245418}.
The corresponding charging time has also been conjectured and numerically validated.
Despite these results, analytical results regarding the conditions for optimal energy storage and the determination of charging time in the central-spin model remain lacking. 
In particular, can we provide rigorous proof for the conditions and charging time for optimal energy storage in the non-TC limit? 
Additionally, even for the TC limit, different initial configurations in the central-spin model can be used to realize it. 
Do these various configurations all guarantee optimal energy storage? 
Do they yield the same charging time?
Addressing these questions will help distinguish the central-spin model from its behavior under the TC limit and deepen our understanding of the nonequilibrium dynamics in the central-spin model.

In this paper we address these questions systematically.
In Sec. II we introduce the central-spin model and the invariant subspace approach used for the subsequent analysis. 
The quantities that characterize optimal energy storage and the collective charging advantage are also given.
Section III employs the HP transformation to identify four initial configurations that achieve the TC limits. 
For the first three cases, we rigorously demonstrate the emergent SU(2) symmetry that ensures optimal energy storage and analytically determine the corresponding charging time. 
Based on these findings, we prove that the collective charging advantage scales with $N_b$ in the first fully charged TC limit. 
For the fourth case, numerical calculations indicate the absence of optimal energy storage.
In Sec. IV we explore the non-TC limits. 
We analytically demonstrate a novel scenario for achieving optimal energy storage while maintaining an $N_b$ scaling for the collective charging advantage.
Additionally, numerical analysis reveals another non-TC configuration that attains asymptotic optimal energy storage, with a collective charging advantage scaling as $N_b^{0.8264}$.
The origin of the collective charging advantage is discussed in Sec. V using the quantum speed limit and a multipartite entanglement witness.
A summary is given  in Sec. VI.

\section{Central-spin quantum battery }
The central-spin QB is described by the  Hamiltonian
\begin{eqnarray}\label{CSB}
 H&=& H_{b}+H_{c}+ H_{I},\n
H_{b}&=&\omega_b S^z, \n
H_{c}&=&\omega_c J^z, \n
H_{I}&=&A(S^+J^-+S^-J^+),
\end{eqnarray}
where $H_b$, $H_c$, and $H_I$ correspond to the Hamiltonian of the
battery, charger, and their homogeneous flip-flop spin interactions in the $x$-$y$ plane, respectively.
Here, $S^{\alpha}\!=\!\sum_{j=1}^{N_b}\sigma_j^\alpha/2$ and $J^{\alpha}\!=\!\sum_{k=1}^{N_c}\sigma_k^\alpha/2$ with $\alpha\!=\!x,y,z$ are the collective spin operators for the $N_b$ spin-1/2 battery cells and $N_c$ spin-1/2 charger units, respectively. 
The spin ladder operators are given by $J^\pm\!=\!J^x\pm iJ^y$ and $S^\pm\!=\!S^x\pm iS^y$.

Initially, the battery is set to the ground state of $H_b$ with $N_b$ spins down, namely, $|0\rangle_{b}\!\equiv|\!\downarrow_{1},\downarrow_{2},\cdot\cdot\cdot,\downarrow_{N_{b}}\rangle_{b}$, while the charger is prepared in a high-energy Dicke state $|m\rangle_{c}$ with $m$ spins up [see Fig.\ref{fig0}(a)]. 
The charging dynamics is performed by turning on the interactions $H_I$.
To ensure that the energy injected into the battery exclusively comes from the charger,  we require $[H_b+H_c,H]\!=\!0$ and thus we only consider the resonant case with $\omega_b\!=\!\omega_c\!\equiv\!\omega$.
Therefore, the free part in the Hamiltonian~(\ref{CSB}) is a constant and we will ignore it.
In terms of the invariant subspace  $\mathcal H_{m}=\{|0\rangle_{b}|m\rangle_{c},|1\rangle_{b}|m\!-\!1\rangle_{c},\cdot\cdot\cdot,|d\rangle_{b}|m\!-\!d\rangle_{c}\}$ containing the initial state $\ket{\psi(0)}\!=\!\ket{0}_b\ket{m}_c$,  the  Hamiltonian $H$ defined in Eq.~(\ref{CSB}) can be represented as a $(d+1)\times(d+1)$ matrix in the invariant subspace $\mathcal H_{m}$ by using the Dicke basis 
        \begin{equation}\label{N-1}
                \bm {H}=\begin{pmatrix}
                   0 & u_1    &                           \\
                    u_1 & 0    & u_2     &                 \\
                        & \ddots & \ddots  & \ddots  &       \\
                        &        & u_{d-1} & 0 & u_{d} \\
                        &        &         & u_{d}   & 0
                  \end{pmatrix},
        \end{equation}
where we have shifted the ground-state energy to cancel the  diagonal terms.
The off-diagonal terms are given by
\begin{eqnarray}
&&u_{j}(N_b,N_c,m)\n 
& &= A\sqrt{j(N_b-j+1)(N_c-m+j)(m-j+1)}.
\end{eqnarray}
The dimension of the effective Hamiltonian~(\ref{N-1}) is $d\!+\!1$ with $d\equiv\min\{{N_{b},m}\}$, which scales linearly with the minimal number of battery cells or the charger units.
Then it is easy to diagonalize $\bm{H}$ to calculate the dynamics 
\begin{eqnarray}\label{WaveFunc}
\bm{\psi}(t)=e^{-i\bm Ht}(1\ 0\ \ldots\ 0)^T.
\end{eqnarray}
Thus, the evolution of the total quantum state is expressed as
\begin{eqnarray}\label{total-state}
\ket{\psi(t)}=\bm\psi_0(t)\ket{0}_b\ket{m}_c+\cdots+\bm\psi_{d}(t)\ket{d}_b\ket{m-d}_c.\n
\end{eqnarray}
The reduced density matrix of the battery state is then obtained as
\begin{eqnarray}\label{RDM}
\rho_b(t)&\equiv& \tr_c[\ket{\psi(t)}\bra{\psi(t)}]\n
&=&|\bm \psi_0(t)|^2\ket{0}\bra{0}+\cdots+|\bm\psi_{d}(t)|^2\ket{d}\bra{d}.
\end{eqnarray}

The energy transferred from the charger to the battery is given by
\begin{eqnarray}\label{Delta-E}
\Delta E(t)=\ex{\psi(t)|H_b|\psi(t)}-\ex{\psi(0)|H_b|\psi(0)},
\end{eqnarray}
where $H_b$ denotes the Hamiltonian of the battery.
The charging time, denoted by $T$, is defined as the time required to achieve the maximum value of $\Delta E(t)$.
At the charging time $T$, the charger has injected as much energy as possible into the battery.
However, optimal energy storage occurs only in the following two cases: (i) The battery absorbs all the energy initially stored in the charger or (ii) the battery absorbs some energy from the charger and reaches a fully charged state $\ket{\up\cdots\up}$.
To quantify the performance of stored energy, we define $\eta(t)$ as
\begin{eqnarray}\label{eta}
\eta(t)\equiv \frac{\Delta E(t)}{\ex{\psi_{\rm opt}|H_b|\psi_{\rm opt}}-\ex{\psi(0)|H_b|\psi(0)}},
\end{eqnarray}
where the denominator represents the optimal energy storage, given by $\omega\min\{N_b,m\}$, and  $\ket{\psi_{\rm opt}}\!=\!\ket{\min\{N_b,m\}}_b\ket{m\!-\!\min\{N_b,m\}}_c$ is the optimal charged state.  
Therefore, the condition for optimal charging can be expressed as $\eta(T)\!=\!1$.
One of our main goals is to analytically determine which types of initial states can achieve optimal energy storage.

Under the restriction of optimal energy storage, our second goal is to demonstrate the collective charging advantage analytically. 
In the collective charging scheme, shown in Fig.\ref{fig0}(b), $N_b$ battery cells collectively interact with  $N_c$ charging units, with the charger initially having $m$ spins in the excited state.
The collective charging power is defined as 
\begin{eqnarray}
P_{\#}(N_b,N_c,m)\equiv\frac{\Delta E(T)}{T(N_b,N_c,m)},
\end{eqnarray}
which depends solely on the battery size $N_b$, the charger size $N_c$, and the initial number of excited charger spins $m$.
For a fair comparison, the resources used in the parallel charging scheme [see Fig.~\ref{fig0}(c)] must be equivalent to those used in the collective charging scheme. 
In the parallel charging scheme, there are $N_b$ individual single-cell QBs, where each battery cell interacts with $ N_c/N_b$ charger units, with $ m/N_b$  spins initially excited.
To make sense for this setting, we require that $ N_c\!\geq\!N_b$ and $ m\!\geq\!N_b$, and we will define the collective charging advantage (\ref{Charing-Adv}) only for this case.
According to Ref.~\cite{PhysRevB.104.245418}, the charging power for a single-cell central-spin QB is given by  
\begin{eqnarray}\label{P-single}
P_{\rm single-cell}&=&\frac{2\omega}{\pi } u_{1}(1, N_c/N_b,m/N_b)\n
&=&\frac{2\omega A}{\pi } \sqrt{\frac{m}{N_b}\left(\frac{N_c}{N_b}-\frac{m}{N_b}+1\right)}.
\end{eqnarray}
In the parallel charging scheme, the charging time remains the same as in the single-cell QB case, but the total stored energy will be $N_b$ times larger. 
Therefore, the parallel charging power is given by 
\begin{eqnarray}
P_{\parallel}\equiv N_b P_{\rm single-cell}.
\end{eqnarray} 
The collective charging advantage is thus quantified by 
\begin{eqnarray}\label{Charing-Adv}
\Gamma=\frac{P_{\#}}{P_{\parallel}}=\frac{P_{\#}}{N_b P_{\rm single-cell}}.
\end{eqnarray}
If $\Gamma\!\sim\! N_b^\alpha$ with $\alpha\!>\!0$, we say the collective charging advantage exists.

\begin{figure*}[t]
        \includegraphics[width=1.7\columnwidth]{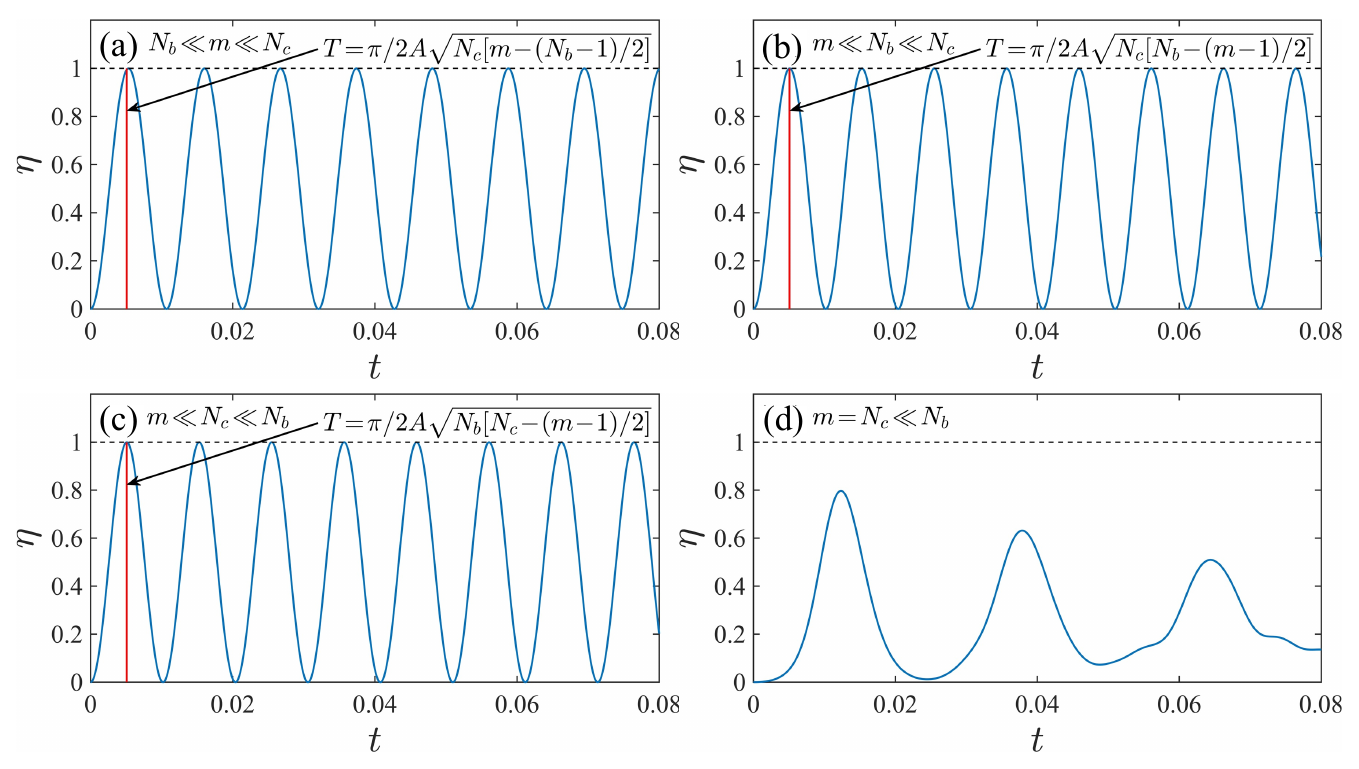}  \caption{
        Charging dynamics for four TC limits: (a) $N_b\!\ll\!m\!\ll\!N_c$; (b) $m\!\ll\!N_b\!\ll\!N_c$; (c) $m\!\ll\!N_c\!\ll\!N_b$; and  (d) $m\!=\!N_c\!\ll\!N_b$. The blue lines are numerically obtained and the red lines are theoretical results. The coupling strength is set to $A\!=\!1$.}\label{fig1}
    \end{figure*}

\section{Optimal energy storage and collective charging speedup at TC Limits}
In previous work~\cite{PhysRevB.104.245418} we analytically demonstrated a case of optimal energy storage for  $2\!=\!N_b\!\ll\! m\!\ll\! N_c$ and numerically verified the case of  $N_b\!\ll\! m\!\ll\! N_c$ with arbitrary $N_b$ to achieve optimal energy storage. 
However, the corresponding charging time was not determined.
Actually, under these initial conditions, the central-spin model can be mapped to the TC model, which we refer to as the TC limit. 
In this section we extend our previous work to find more TC limits and then we  use the theory developed in Ref.~\cite{PhysRevA.109.012204} to determine not only the possibility for achieving optimal energy storage but also the collective charging advantage.
We emphasize that the emergent SU(2) symmetry in the dynamics plays a crucial role in achieving optimal energy storage and is key to analytically calculating the collective charging advantage.

For the case of $N_b\!\ll\!m\!\ll\!N_c$, due to the conservation of the total excitation number $[J^z\!+\!S^z,H]\!=\!0$, the number of excitations in the charger part cannot exceed the initial value  $m$.
Since $\lim_{N_c\to\infty} m/N_c=0$, the charger remains close to the vacuum state $\ket{0}_c$ throughout the dynamics at the limit $N_c\!\to\!\infty$ .
Therefore, we can apply the  HP transformation to the charger Hamiltonian as 
\begin{eqnarray}\label{HP}
&&J^+\to \sqrt{N_c} a^\dag \sqrt{1-\frac{a^\dag a}{N_c}},\n 
&&J^z\to -\frac{N_c}{2}+a^\dag a,
\end{eqnarray}
where  $\sqrt{1-a^\dag a/N_c}\simeq 1$ because $\ex{\psi(t)|a^\dag a|\psi(t)}/N_c\!=\!\ex{\psi(t)|J^z|\psi(t)}/N_c\!+\!1/2\!\leq\!m/N_c\!\to\! 0$.
Under this approximation, the original Hamiltonian~(\ref{CSB}) reduces to
\begin{eqnarray}\label{TC-1}
H\overset{N_b\!\ll\!m\!\ll\!N_c}{\longrightarrow} A\sqrt{N_c} (S^+ a +S^- a^\dag),
\end{eqnarray} 
where $a$ and $a^\dag$ are bosonic annihilation and creation operators, respectively.
In this correspondence,  the number of charger spin excitations $m$ in the initial state $\ket{\psi(0)}\!=\!\ket{0}_b\ket{m}_c$ corresponds to the photon number.
In Ref.~\cite{PhysRevA.109.012204} we have shown that for TC battery~(\ref{TC-1}) and initial state $\ket{\psi_{\rm TC}(0)}=\ket{0}_b\ket{n_0}_c$, with $\ket{n_0}$ being a Fock state with $n_0$ photons,   a distinct SU(2) symmetry  emerges from the dynamics for the limiting cases $n_0 \!\gg\! N_b$ or $N_b \!\gg\! n_0$.
This results in optimal energy storage given by $\Delta E(T_{\rm TC})\!=\!\omega\min\{N_b,n_0\} $ with the charging time 
\begin{eqnarray}\label{Yang-TC}
T_{\rm TC}=\frac{\pi}{2A\sqrt{N_c}\sqrt{\max\{N_b,n_0\}-\frac{\min\{N_b,n_0\}-1}{2}}}.
\end{eqnarray}
Therefore, we conclude that optimal energy storage in the central-spin QB can be achieved with the initial state  $\ket{\psi(0)}=\ket{0}_b\ket{m}_c$, with $N_b\!\ll\!m\!\ll\!N_c$ [see Fig.\ref{fig1}(a)].
By substituting $n_0\!=\!m$ into Eq.~(\ref{Yang-TC}), the charging time is given by
\begin{eqnarray}\label{T-TC-1}
T(N_b\!\ll\!m\!\ll\!N_c)=\frac{\pi}{2A\sqrt{N_c}\sqrt{m-\frac{N_b-1}{2}}},
\end{eqnarray}
which is numerically verified in Fig.~\ref{fig1}(a).
Then by substituting Eqs.~(\ref{P-single}) and (\ref{T-TC-1}) into Eq.~(\ref{Charing-Adv}), the collective charging advantage is obtained as  
\begin{eqnarray}
\Gamma(N_b\!\ll\!m\!\ll\!N_c)=
\frac{\sqrt{N_c}\sqrt{m-\frac{N_b-1}{2}}}{ \sqrt{\frac{m}{N_b}\left(\frac{N_c}{N_b}-\frac{m}{N_b}+1\right)}}\simeq N_b,
\end{eqnarray} 
which  indicates an $N_b$-fold collective charging speedup compared to the parallel charging scheme.

It is important to emphasize that the TC Hamiltonian~(\ref{TC-1}), which describes the central-spin QB, only requires the number of excited charger spins to be much smaller than the total number of charger spins, i.e., $m\!\ll\! N_c$.
Moreover, the condition $n_0\!\ll\! N_b$ for optimal energy storage in the TC battery implies another scenario $m\!\ll\! N_b\!\ll\! N_c$, which also facilitates optimal energy storage in the central-spin QB [see Fig.~\ref{fig1}(b)].
By setting $n_0\!=\!m$ in Eq.~(\ref{Yang-TC}), we obtain the charging time as  
\begin{eqnarray}
T(m\!\ll\!N_b\!\ll\!N_c)=\frac{\pi}{2A\sqrt{N_c}\sqrt{N_b-\frac{m-1}{2}}},
\end{eqnarray} 
which is numerically verified in Fig.~\ref{fig1}(b).
Note that calculating the charging advantage in this case makes no sense because  $m/N_b\!\to\! 0$.
Therefore, we will only consider the charging advantage for $N_b\!\leq\! m$ and $N_b\!\leq\! N_c$   in the following discussion.

Unlike the two situations discussed above, we can also employ the HP transformation to map the battery onto a bosonic mode, facilitating the construction of a TC QB from the central-spin  QB. 
Now let us consider the case where $m\!\ll\! N_c\!\ll\! N_b$.
In this regime, the possible maximum excitation for the battery part is 
$m$, which is significantly smaller than the battery size $N_b$. 
Consequently, the HP transformation can be applied to the battery spins, yielding
\begin{eqnarray}
&&S^+\to \sqrt{N_b} b^\dag \sqrt{1-\frac{b^\dag b}{N_b}},\n 
&&S^z\to -\frac{N_b}{2}+b^\dag b,
\end{eqnarray}
where, by approximating   $\sqrt{1-b^\dag b/N_b}\simeq 1$, the central-spin QB in Eq.~(\ref{CSB}) reduces to the TC QB,
\begin{eqnarray}\label{TC-2}
H\overset{m\!\ll\!N_c\!\ll\!N_b}{\longrightarrow} A\sqrt{N_b} (b J^+ +b^\dag J^-),
\end{eqnarray} 
where the battery has become a photon field.
From the perspective of matrix representation in the invariant subspace [Eq.~(\ref{N-1})], the above HP transformation approximates the matrix element  $u_j$ as 
\begin{eqnarray}\label{u_j-1}
u_j&=& A\sqrt{j(N_b-j+1)(N_c-m+j)(m-j+1)}\n 
&\overset{m\!\ll\! N_b}{\longrightarrow}&A\sqrt{N_b}\sqrt{j(N_c-m+j)(m-j+1)}
\end{eqnarray}
with $j\!=\!1,\ldots,m$.
Considering the additional condition $m\ll N_c$, we can further approximate Eq.~(\ref{u_j-1}) as
\begin{eqnarray}\label{u_j-2}
u_j 
&\overset{m\!\ll\! N_c\!\ll\!N_b}{\longrightarrow}&A\sqrt{N_b}\sqrt{N_c-\frac{m-1}{2}}\sqrt{j(m-j+1)},
\end{eqnarray}
which implies that the effective Hamiltonian~(\ref{N-1}) is exactly the generator of the SU(2) algebra
\begin{eqnarray}
 \bm {H}\overset{m\!\ll\! N_c\!\ll\!N_b}{\longrightarrow} \sqrt{N_b}\Omega \bm {L^x},
\end{eqnarray} 
where $\Omega\equiv 2A\sqrt{N_c-(m-1)/2}$ is the generalized Rabi frequency and $\bm{L^x}$ is the $x$-direction generator of SU(2) algebra in the spin-$m$ representation.
The evolution of energy storage~(\ref{Delta-E}) can be represented as 
\begin{eqnarray}
\Delta E(t)=\frac{m\omega}{2} +\bm{\psi}^\dag(0)\bm{L^z}(t)\bm{\psi}(0)\omega,
\end{eqnarray}
where  $\bm{L^z}(t)\!=\!\exp(it\Omega\sqrt{N_b}\bm{L^x})\bm{L^z}\exp(-it\Omega\sqrt{N_b}\bm{L^x})$ and $\bm{\psi}(0)\!=\!(1,0,\cdots,0)^T$.
Using the Baker-Campbell-Hausdorff formula, we have
\begin{eqnarray}
\bm{L^z}(t)=\sin(\Omega\sqrt{N_b}t)\bm{L^y}+\cos(\Omega\sqrt{N_b}t)\bm{L^z},
\end{eqnarray}
and thus
\begin{eqnarray}
\Delta E(t)=\frac{m\omega}{2}[1-\cos(\Omega \sqrt{N_b}t)],
\end{eqnarray}
which implies that optimal energy storage can be achieved at the charging time 
\begin{eqnarray}
&&T(m\!\ll\! N_c\!\ll\!N_b)=\frac{\pi}{\Omega\sqrt{N_b}}\n 
& &=\frac{\pi}{2A\sqrt{N_b}\sqrt{N_c-\frac{m-1}{2}}}
\end{eqnarray}
 [see Fig.\ref{fig1}(c) for numerical verification].

However, we claim that optimal energy storage cannot be achieved in all TC limits, where the emergence of SU(2) symmetry plays a crucial role. As a specific case, we consider the scenario where  $m\!=\!N_c\!\ll\! N_b$.
Given the ratio  $m/N_b\to0$,  the maximum possible number of excitations in the battery during the dynamics,  i.e., $m$, is close to zero.
Consequently, the effective Hamiltonian remains identical to Eq.(\ref{TC-2}). Nevertheless, since  $m=N_c$ we are unable to apply the same approximation used in deriving  Eq.~(\ref{u_j-2}).
As a result, it is not possible to obtain an SU(2) generator to describe the charging dynamics in this particular TC limit. 
The absence of SU(2) symmetry in the regime  $m\!=\!N_c\!\ll\! N_b$ therefore prevents the realization of optimal energy storage, as shown in Fig.~\ref{fig1}(d).

    \begin{figure}[t]
        \includegraphics[width=\columnwidth]{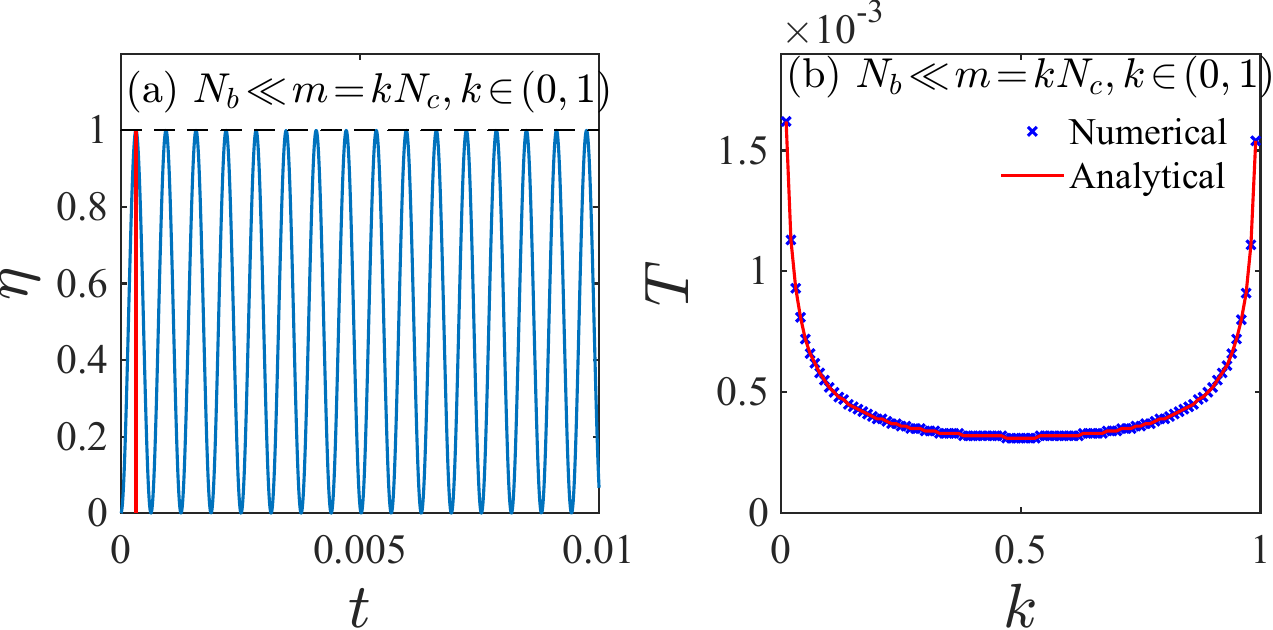}\caption{ (a) Charging dynamics in the non-TC limit: $N_b\!\ll\!m\!=\!kN_c$ with $0\!<\!k\!<\!1$. 
        The blue lines represent numerical results, while the red lines correspond to theoretical predictions. (b) The analytical charging time (\ref{T-k}) as a function of $k$ is numerically verified for $N_b\!\ll\! m\!=\!kN_c$ with $0\!<\!k\!<\!1$. The coupling strength is set to $A\!=\!1$.}\label{fig2}
    \end{figure}

\section{Optimal energy storage and collective charging speedup  at non-TC Limits}
Having successfully applied the SU(2) symmetry theory to address the optimal energy storage problem in the TC limits, in this section we use it again to investigate the problem in the non-TC limits.
The non-TC limits distinguish the central-spin QB from the TC QB, representing a unique regime specific to the central-spin QB.

The first non-TC limit considered here is $N_b\!\ll\!m\!=\!kN_c$, with $k\!\in\!(0,1)$.
Since $N_b\ll m$, the charger part can be considered essentially unchanged during the dynamics.
Therefore, macroscopic excitations are always present in the charger, i.e., 
\begin{eqnarray}
&&\lim_{N_c\to\infty}\frac{\ex{\psi(t)|\hat a^\dag a|\psi(t)}}{N_c}\n
& &=\lim_{N_c\to\infty}\frac{\ex{\psi(t)|J^z|\psi(t)}}{N_c}+\frac{1}{2}=\frac{m}{N_c}=k,
\end{eqnarray}
which prevents the direct application of the HP transformation given in Eq.~(\ref{HP}).
To circumvent this issue, we can first apply the HP transformation (\ref{HP}) and then shift the bosonic operators as 
\begin{eqnarray}\label{shift}
\hat a=\hat c+\sqrt{kN_c}.
\end{eqnarray}
With this shift, we obtain
\begin{eqnarray}\label{condition}
&&\lim_{N_c\to\infty}\frac{\ex{\psi(t)|\hat c^\dag c|\psi(t)}}{N_c}\n 
& &=\lim_{N_c\to\infty}\frac{\ex{\psi(t)|\hat a^\dag a|\psi(t)}}{N_c}-\frac{kN_c}{N_c}=0.
\end{eqnarray}

   \begin{figure}[t]
        \includegraphics[width=\columnwidth]{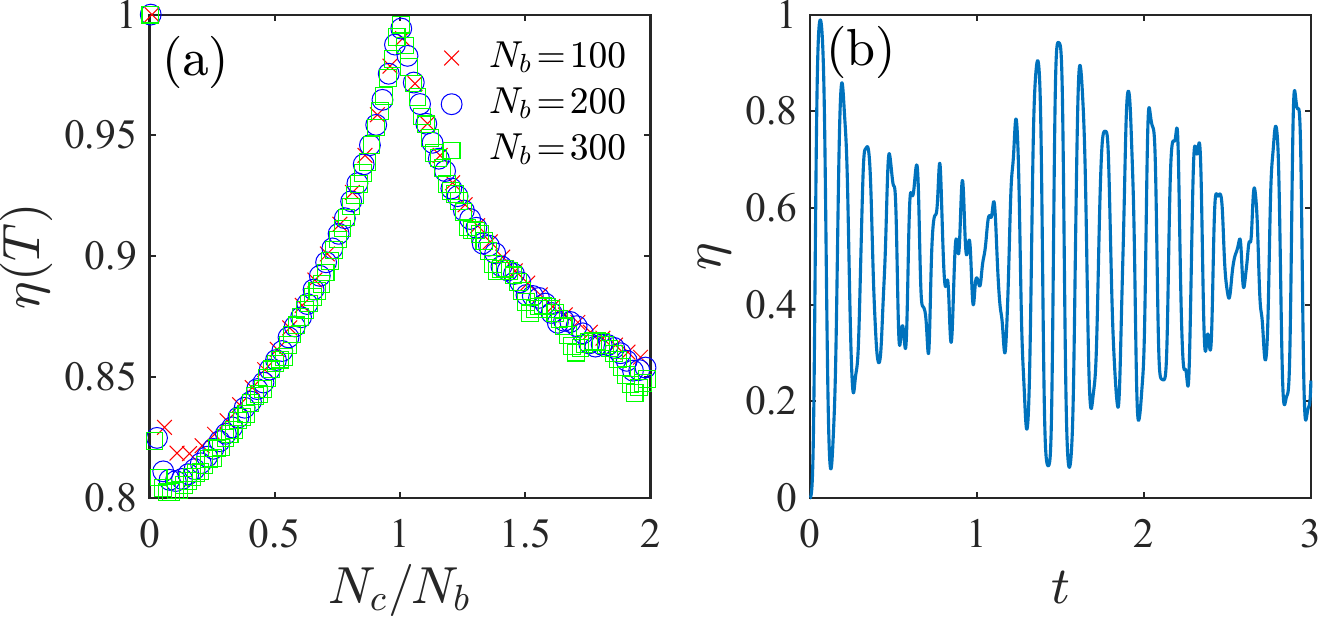}\caption{(a) Maximum  energy storage efficiency $\eta(T)$ as a function of $N_c/N_b$ for the case $m\!=\!N_b$. (b) Charging dynamics for $m\!=\!N_b\!=\!N_c=100$. The coupling strength is set to $A\!=\!1$.}\label{fig3}
    \end{figure}

Substituting Eq.~(\ref{shift}) into Eq.~(\ref{HP}), we have 
\begin{eqnarray}\label{Shift-HP}
J^+&=&\sqrt{(1-k)N_c}(c^\dag+\sqrt{kN_c})\n 
& &\times\sqrt{1-\frac{c^\dag c+\sqrt{kN_c}(c+c^\dag)}{(1-k)N_c}}\n 
&=&\sqrt{k(1-k)}N_c+ O(\sqrt{N_c}),
\end{eqnarray}
where we have used Eq.~(\ref{condition}).
To leading order in  $N_c$, the central-spin model in Eq.~(\ref{CSB}) reduces to
\begin{eqnarray}\label{eff-H-k}
H&\overset{N_b\!\ll\! m=kN_c}{\longrightarrow}& A \sqrt{k(1-k)}N_c(S^-+S^+)\n 
 &=&2A \sqrt{k(1-k)}N_c S^x,
\end{eqnarray} 
which is exactly the generator of the SU(2) algebra and therefore means that SU(2) symmetry also appears in this non-TC limit.
The evolution of energy storage~(\ref{Delta-E}) in this case can be expressed as 
\begin{eqnarray}
\Delta E(t)=\frac{N_b\omega}{2} +\ex{\psi(0)|S^z(t)|\psi(0)}\omega,
\end{eqnarray}
where  $S^z(t)\!=\!\exp(itH)S^z\exp(-itH)$.
Using the Baker-Campbell-Hausdorff formula, we find
\begin{eqnarray}
S^z(t)&=&\sin(2A\sqrt{k(1-k)}N_ct)S^y\n 
& &+\cos(2A\sqrt{k(1-k)}N_ct)S^z.
\end{eqnarray}
Thus, the energy evolution is given by
\begin{eqnarray}
\Delta E(t)=\frac{N_b\omega}{2}[1-\cos(2A\sqrt{k(1-k)}N_ct)],
\end{eqnarray}
which indicates that optimal energy storage is achieved at the charging time
\begin{eqnarray}\label{T-k}
T(N_b\!\ll\!m\!=\!kN_c)=\frac{\pi}{2AN_c\sqrt{k(1-k)}},
\end{eqnarray}
as confirmed numerically in Fig.\ref{fig2}(b).
Verification of optimal energy storage is shown in Fig. \ref{fig2}(a).
By substituting  Eq.~(\ref{T-k}) into Eq.~(\ref{Charing-Adv}), we obtain the collective charging advantage as
\begin{eqnarray}
\Gamma(N_b\!\ll\!m\!=\!kN_c)=
\frac{N_c\sqrt{k(1-k)}}{ \sqrt{\frac{m}{N_b}\left(\frac{N_c}{N_b}-\frac{m}{N_b}+1\right)}}\simeq N_b.
\end{eqnarray}
In summary, we show that SU(2) symmetry can also emerge in the non-TC limit, leading to optimal energy storage with a collective charging advantage scaling as $N_b$.

   \begin{figure}[t]
        \includegraphics[width=0.7\columnwidth]{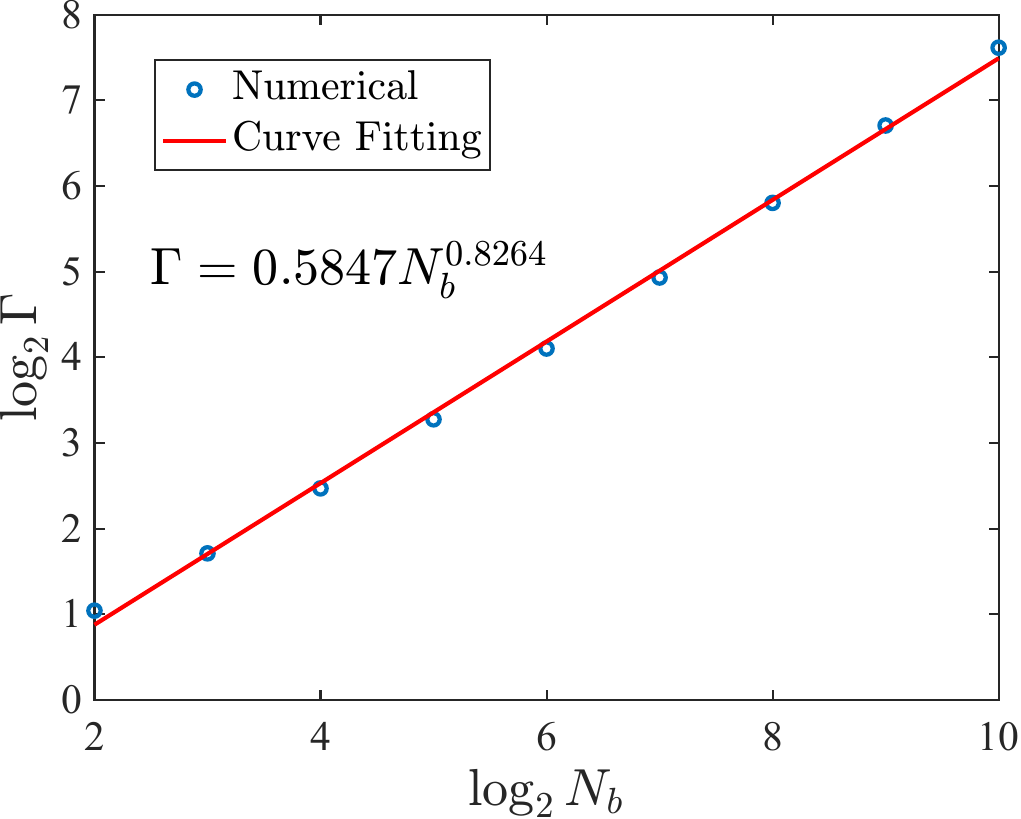}\caption{Scaling of the collective charging advantage $\Gamma$ as a function of the number of battery cells  $N_b$ in the case of $N_b\!=\!m\!=\!N_c$. The coupling strength is set to $A\!=\!1$.}\label{fig4}
    \end{figure}

Note that Eq.~(\ref{T-k}) does not hold for the case $k\!=\!1$, or equivalently $m\!=\!N_c$, because the approximation of $J^+$ is incorrect in this scenario, as the zeroth-order term vanishes.
To investigate the region where $m\!=\!N_c$, we perform numerical calculations (see Fig. \ref{fig3}).
In Fig.~\ref{fig3}(a) we obtain the maximum energy transferred into the battery and plot the energy storage efficiency $\eta(T)$, defined in Eq.~(\ref{eta}), at the charging time $T$ for various values of $N_b$.
Surprisingly, $\eta(T)$ exhibits a universal behavior, with all curves collapsing onto a single curve after rescaling.
Moreover, we observe that $\eta(T)\!\to\! 1$ for the case $N_c\!=\!N_b$, indicating the possibility of achieving optimal energy storage.
In Fig.~\ref{fig3}(b) we plot the charging dynamics for $N_b\!=\!m\!=\!N_c$, which shows asymptotic optimal energy storage, meaning that the first maximum of $\eta$ approaches one.
However, due to the absence of dynamically emergent SU(2) symmetry, this case differs significantly from the previous situations and does not exhibit sine like oscillations. 
In Fig.~\ref{fig4} we plot the collective charging advantage for  $N_b\!=\!m\!=\!N_c$ and find that $\Gamma\propto N_b^{0.8264}$.
Although $\Gamma$ does not scale exactly with $N_b$ in this case, it still demonstrates a collective charging advantage close to an $N_b$-scaling behavior.

\section{Origin of collective charging advantage}
We emphasize that there are two distinct origins of the collective charging advantage: the classical origin~\cite{PhysRevB.99.205437,zhang2023enhanced,RevModPhys.96.031001} and the quantum origin~\cite{PhysRevLett.125.236402}.
The classical origin arises from the scaling amplification of the norm of the collective charging Hamiltonian, rather than from quantum correlations, such as multipartite entanglement.
We argue that the collective charging advantage observed in the central-spin quantum battery is also classical in nature, similar to that of the Dicke battery~\cite{PhysRevLett.120.117702}.

The first piece of evidence comes from the quantum speed limit~\cite{mandelstam1945uncertainty,deffner2017quantum}
\begin{eqnarray}
T\geq \frac{\pi\hbar}{2\Delta E},
\end{eqnarray}
where $T$ is the charging time, $\hbar\!=\!1$ in all calculations, and $\Delta E\!=\!\sqrt{\ex{\psi(0)|H^2|\psi(0)}-\ex{\psi(0)|H|\psi(0)}^2}$.
If the charging time saturates the quantum speed limit, then the charging time $T'$ generated by the dynamics $H'\!=\!N_bH$ will be $1/N_b$ times the original charging time $T$, which is a purely classical effect.
Therefore, we introduce the charging rate defined as 
\begin{eqnarray}\label{gamma}
\gamma\equiv \frac{\tau_{\rm QSL}}{T}
\end{eqnarray}
to witness the classical origin of the collective charging advantage, where $\tau_{\rm QSL}=\pi/2\Delta E$.

A typical parallel charging scheme is governed by Hamiltonian $H_{\parallel}\!=\!S^x\!=\!\sum_{j=1}^{N_b}\sigma_j^x/2$.
It follows that $T_{\parallel}\!=\!\pi$ and $\Delta E_{\parallel}=\sqrt{N_b}/2$.
Thus, the typical charging rate for a parallel charging scheme is 
\begin{eqnarray}\label{gamma-p}
\gamma_{\parallel}=\frac{1}{\sqrt{N_b}}.
\end{eqnarray}
For the collective charging scheme in the central-spin QB, the variance of charging Hamiltonian~(\ref{CSB}) is given by 
\begin{eqnarray}
\Delta E&=&A\sqrt{\ex{0|S^-S^+|0}\ex{m|J^+J^-|m}}\n 
&=&A\sqrt{N_bm(N_c-m+1)}.
\end{eqnarray}
Now we consider the charging rate for two cases with collective charging advantage.
The first case corresponds to the TC limit discussed in Sec. III, i.e., $N_b\!\ll\!m\!\ll\!N_c$. 
Substituting the charing time~(\ref{T-TC-1}) into the definition of charging rate (\ref{gamma}), we obtain 
\begin{eqnarray}\label{gamma-1}
\gamma(N_b\!\ll\!m\!\ll\!N_c)\!=\!\frac{\sqrt{N_c}\sqrt{m-\frac{N_b-1}{2}}}{\sqrt{N_bm(N_c-m+1)}}\simeq\frac{1}{\sqrt{N_b}}.
\end{eqnarray}  
The second case corresponds to the non-TC limit $N_b\!\ll\!m\!=\!kN_c$ discussed in Sec. IV. 
By substituting the charing time~(\ref{T-k}) into the charging rate (\ref{gamma}), we have
\begin{eqnarray}\label{gamma-2}
\gamma(N_b\!\ll\!m\!=\!kN_c)\!=\!\frac{N_c\sqrt{k(1-k)}}{\sqrt{N_bm(N_c-m+1)}}\!\simeq\! \frac{1}{\sqrt{N_b}}.
\end{eqnarray}
Since the collective charging rate in the central-spin quantum battery, given by Eqs.~(\ref{gamma-1}) and (\ref{gamma-2}), is the same as the parallel charging rate in Eq.~(\ref{gamma-p}), we conclude that the collective charging advantage in the central-spin quantum battery is also classical due to the super linear scaling of the bandwidth of the charging Hamiltonian.

   \begin{figure}[t]
        \includegraphics[width=0.75\columnwidth]{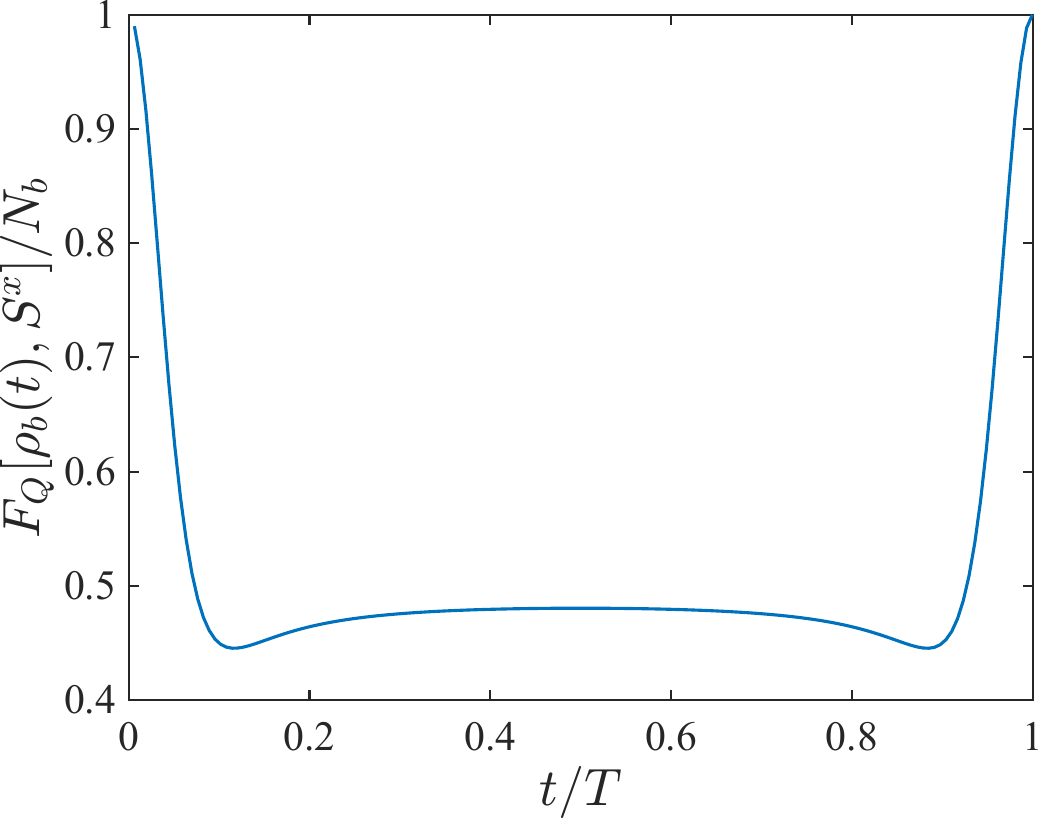}\caption{
Evolution of the rescaled QFI for the battery state in the cases of $N_b\!\ll\!m\!\ll\!N_c$ and $N_b\!\ll\!m\!=\!kN_c$, in which an $N_b$-fold enhancement in speedup exists. 
The number of battery cells is set to $N_b\!=\!50$.}\label{figQFI}
    \end{figure}

\begin{table*}[t]
\centering
\caption{Optimal energy storage, collective charging advantage, and emergent SU(2) symmetry in the central-spin QB.}
\begin{tabular}{c|c|c|c|c}
\hline\hline
 Case& Charging Time & Optimal Energy Storage & Emergence SU(2) Symmetry & Collective Charging Advantage  \\
 \hline
 $N_b\!\ll\!m\!\ll\!N_c$& $\frac{\pi}{2A\sqrt{N_c}\sqrt{m-\frac{N_b-1}{2}}}$ & Yes & Yes & $N_b$ \\
  \hline
 $m\!\ll\!N_b\!\ll\!N_c$& $\frac{\pi}{2A\sqrt{N_c}\sqrt{N_b-\frac{m-1}{2}}}$ & Yes & Yes & NA \\
  \hline
 $m\!\ll\!N_c\!\ll\!N_b$& $\frac{\pi}{2A\sqrt{N_b}\sqrt{N_c-\frac{m-1}{2}}}$ & Yes & Yes & NA \\
  \hline
 $m\!=\!N_c\!\ll\!N_b$ & Numerical  & No & No & NA \\
  \hline
$N_b\!\ll\!m\!=\!kN_c$\\ $k\!\in\!(0,1)$ & $\frac{\pi}{2AN_c\sqrt{k(1-k)}}$ & Yes & Yes & $N_b$ \\
  \hline
  $m\!=\!N_c\!=\!N_b$& Numerical & Asymptotically & No & $N_b^{0.8264}$ \\
\hline\hline
\end{tabular}\label{Table}
\end{table*}

The second piece of evidence comes from analyzing the multipartite entanglement structure during the charging dynamics.
To witness multipartite entanglement, we adopt the quantum Fisher information (QFI).
For a generic $N$-qubit state $\rho\!=\! \sum_{\lambda} q_{\lambda} \ket{\psi_\lambda} \bra{\psi_\lambda}$ (in diagonal form) and an operator $\hat{\mathcal O}$, the QFI is given by~\cite{PhysRevLett.72.3439}
\begin{eqnarray}\label{QFI-def}
F_Q[\rho, \hat{\mathcal O}]\!=\! 2\sum_{\lambda,\lambda'} \frac{(q_\lambda - q_{\lambda'})^2}{q_\lambda + q_{\lambda'}} \vert \bra{\psi_\lambda} \hat{\mathcal O} \ket{\psi_{\lambda'}} \vert^2.
\end{eqnarray}
Violation of the inequality
\begin{eqnarray}
F_Q[\rho_b(t),S^x]\!\leq\! sk^2+r^2,
\end{eqnarray}
for $\hat{\mathcal O}\!=\!S^{x}$
indicates that the battery state $\rho_b(t)$ contains at least $(k\!+\!1)$-particle entanglement among the $N_b$ battery cells~\cite{PhysRevLett.102.100401,PhysRevA.85.022322,PhysRevA.85.022321}, where $s\!=\!\lfloor{N_b/k}\rfloor$ represents the largest integer less than or equal to $N_b/k$, and $r\!=\!N_b\!-\!sk$.
Therefore, if $F_Q[\rho_b(t),S^x]\!>\!N_b$, we can infer the appearance of an entangled battery state.
Conversely, if $F_Q[\rho_b(t),S^x]\!\leq\!N_b$, we cannot extract useful information about the entanglement structure of $\rho_b(t)$.
Actually, QFI is particularly useful for detecting multipartite entangled states close to the Greenberger–Horne–Zeilinger (GHZ)-like states, as the QFI for the GHZ state is $N_b^2$, larger than $N_b$.

Recall the results obtained in the previous sections: We have shown that the appearance of optimal charging is associated with the emergent SU(2) symmetry.
In particular, for cases with collective charging advantage, the Hamiltonian in Eq.~(\ref{N-1}) can be approximated as a generator of SU(2) algebra, i.e.,
\begin{eqnarray}\label{eff-H-CCA}
H\simeq \nu S^x, 
\end{eqnarray}
where $\nu=2A\sqrt{N_c}\sqrt{m-(N_b-1)/2}$
for the case of $N_b\!\ll\!m\!\ll\!N_c$~\cite{footnote}, while $\nu=2A \sqrt{k(1-k)}N_c$ for the case of $N_b\!\ll\!m\!=\!kN_c$, as seen in Eq.~(\ref{eff-H-k}).
Based on the effective Hamiltonian in Eq.~(\ref{eff-H-CCA}), the evolution of the component of the wave function (\ref{WaveFunc}) is obtained as 
\begin{eqnarray}
\bm\psi_m(t)\!=\!\sqrt{\binom{N_b}{m}}
\left[-i\sin\left(\frac{\nu}{2}t\right)\right]^m
\left[\cos\left(\frac{\nu}{2}t\right)\right]^{N_b-m}.
\end{eqnarray}
Therefore, using Eq.~(\ref{RDM}), the reduced state of the battery is given by
\begin{eqnarray}\label{RDM-cca}
\rho_b(t)
&=&\sum_{m=0}^{N_b} q_m(t) \ket{m}\bra{m},
\end{eqnarray}
with eigenvalues
\begin{eqnarray}\label{qm}
q_m(t)&\!=\!&|\bm\psi_m(t)|^2\n 
&\!=\!&\binom{N_b}{m}\left[\sin\left(\frac{\nu t}{2}\right)\right]^{2m}
\left[\cos\left(\frac{\nu t}{2}\right)\right]^{2N_b-2m}.
\end{eqnarray}
Substituting Eq.~(\ref{RDM-cca}) into the definition of QFI~(\ref{QFI-def}), we obtain 
\begin{eqnarray}\label{QFI-final}
&&F_Q[\rho_b(t),S^x]\n 
& &=\sum_{m=0}^{N_b-1} \frac{[q_m(t) - q_{m+1}(t)]^2}{q_m(t) + q_{m+1}(t)} (N_b-m)(m+1),
\end{eqnarray} 
where $q_m(t)$ is given in Eq.~(\ref{qm}).
Figure~\ref{figQFI} plots the evolution of QFI for the battery state, given by Eq.~(\ref{QFI-final}), which is always less than or equal to $N_b$.
Therefore, a GHZ-like multipartite entanglement structure has not been detected among the battery cells using QFI.
With these two pieces of evidence, we conclude that the collective charging advantage observed in the central-spin quantum battery has no genuine quantum origin.

\section{Conclusion}
In this work we investigated the charging dynamics of a central-spin QB.
The problems of optimal energy storage and collective charging advantage were analytically explored within this model. 
We rigorously demonstrated the realization of optimal energy storage in four different initial state settings: (i) $N_b\!\ll\!m\!\ll\!N_c$, (ii) $m\!\ll\!N_b\!\ll\!N_c$, (iii) $m\!\ll\!N_c\!\ll\!N_b$, and (iv) $N_b\!\ll\!m\!=\!kN_c$ [$k\!\in\!(0,1)$], where $N_b$ denotes the number of battery cells,  $N_c$ represents the number of charger units, and $m$ is the initial number of excited charger units.
The first three cases correspond to the TC limits, while the fourth case represents the non-TC limit.
Our findings reveal that, across all these cases, the emergence of SU(2) symmetry from the charging dynamics is crucial for achieving optimal energy storage. 
This emergent symmetry also provides a theoretical approach to analytically determine the charging time. 
In the fully charged scenarios corresponding to cases (i) and (iv), a collective charging speedup of $N_b$-fold is proven in comparison to the parallel charging scheme.
Furthermore, in the case of $N_b\!=\!m\!=\!N_c$ we identified the possibility of asymptotically achieving optimal energy storage, with numerical calculations indicating a collective charging advantage scaling as approximately $\sim N_b^{0.8264}$.
The main results are summarized in Table~\ref{Table}.
By employing the quantum speed limit and multipartite entanglement witness, we identified the classical origin of the collective charging advantage in the central-spin quantum battery.
Nevertheless, the classical collective charging advantage is also valuable for designing high-performance quantum batteries~\cite{RevModPhys.96.031001, ferraro2024reply}.
The theoretical central-spin quantum battery in Eq.~(\ref{CSB}) can be realized in a strongly interacting dipolar spin ensemble~\cite{zu2021emergent}, where a high-spin NV center can effectively correspond to battery cells, and an ensemble of spin-1/2 P1 centers represent charging units.
The key flip-flop spin interactions $H_I$ in Eq.~(\ref{CSB}) can be realized through the magnetic dipole-dipole interaction between NV and P1 centers.
From a practical viewpoint, the case where $N_b\!=\!m\!=\!N_c$ is the most experimentally favorable for demonstrating both optimal energy storage and the collective charging advantage.

Our work highlights the significance of emergent SU(2) symmetry in analytically addressing the challenges of optimal energy storage and collective charging speedup. 
It extends previous studies on the TC QB~\cite{PhysRevA.109.012204} and the central-spin QB~\cite{PhysRevB.104.245418}. 
We believe the methods used here can be applied to other QB models to uncover further the mechanisms underlying optimal energy storage and collective charging speedup.

\section*{ACKNOWLEDGMENTS}
HLS thanks Xu Zhou for providing valuable experimental comments.
This work was supported by the NSFC (Grants No. 12275215, No. 12305028, and No. 12247103), Shaanxi Fundamental Science Research Project for Mathematics and Physics (Grant No. 22JSZ005) and the Youth Innovation Team of Shaanxi Universities.
HLS was supported by the European Commission through the H2020 QuantERA ERA-NET Cofund in Quantum Technologies project ``MENTA''.

\bibliography{Ref}

\end{document}